\pdfoutput=1
\documentclass[11pt]{article}

\usepackage[]{acl}

\usepackage{times}
\usepackage{latexsym}

\usepackage{graphicx}
\usepackage{booktabs}
\usepackage{tabularx}
\usepackage{amsmath, bm}
\usepackage{multirow, multicol}
\usepackage{tabularx}

\usepackage{colortbl}
\definecolor{mygray}{gray}{.9}

\usepackage{verbatim}

\usepackage[T1]{fontenc}
\usepackage[utf8]{inputenc}

\usepackage{microtype}

\usepackage{inconsolata}

\usepackage{graphicx}

\title{Intermediate Distillation: Data-Efficient Distillation \\from Black-Box LLMs for Information Retrieval}

\author{Zizhong Li \quad Haopeng Zhang \quad Jiawei Zhang \\
  IFM Lab, University of California, Davis \\
  \texttt{\{zzoli, hapzhang, jiwzhang\}@ucdavis.edu} \\
  }

\begin{document}
\maketitle

\begin{abstract}
Recent research has explored distilling knowledge from large language models (LLMs) to optimize retriever models, especially within the retrieval-augmented generation (RAG) framework. 
However, most existing training methods rely on extracting supervision signals from LLMs' weights or their output probabilities, which is not only resource-intensive but also incompatible with black-box LLMs. 
In this paper, we introduce \textit{Intermediate Distillation}, a data-efficient knowledge distillation training scheme that treats LLMs as black boxes and distills their knowledge via an innovative LLM-ranker-retriever pipeline, solely using LLMs' ranking generation as the supervision signal. 
Extensive experiments demonstrate that our proposed method can significantly improve the performance of retriever models with only 1,000 training instances. 
Moreover, our distilled retriever model significantly boosts performance in question-answering tasks within the RAG framework, demonstrating the potential of LLMs to economically and effectively train smaller models.
\end{abstract}

% In this paper, we propose an alternative, more efficient method for leveraging the supervision capabilities of LLMs without direct access to their weights. We introduce a multi-step training technique that employs natural language outputs from LLMs as supervision signals. Our experimental results indicate that this method not only improves the performance of information retrievers more than traditional signals but also demonstrates significant advancements in question-answering tasks within the Retrieval-Augmented Generation (RAG) framework. These findings underscore the potential of LLMs to enhance smaller models' training effectively and economically.

\section{Introduction}

The rapid growth and superior performance of large language models (LLMs) \cite{ouyang2022training, openai2023gpt4, wang2024survey} have made them a preferred choice for a wide range of NLP applications \cite{xi2023rise, wang2024survey, wu2023bloomberggpt, zhang2023extractive, zhang2023summit}. 
LLMs have demonstrated robust zero-shot ranking abilities in English and various low-resource languages \cite{adeyemi2023zero, sun2023chatgpt}. Consequently, researchers have applied LLMs to the task of information retrieval, where they outperform previous text search and similarity measurement methods \cite{ma2023fine, xu2024bmretriever}.
% For instance, studies have shown that well fine-tuned LLMs can establish state-of-the-art retrieval systems \cite{ma2023fine, xu2024bmretriever}.
% Meanwhile, LLMs also demonstrate their robust zero-shot ranking capabilities in English documents and various low-resource languages \cite{adeyemi2023zero, sun2023chatgpt}.
% For example, prior studies have found that using a fine-tuned LLM to serve as the retriever and ranker could establish an effective and SOTA retrieval system that outperforms conventional smaller models \cite{ma2023fine, xu2024bmretriever}.
% Moreover, for zero-shot capabilities, LLMs also prove their effectiveness in ranking English documents, as well as some low-resource languages \cite{adeyemi2023zero, sun2023chatgpt}.
% retriever models' performance through joint distillation training methods, thereby further improves LLMs' ability to address this area's task, such as question-answering \cite{kim2024sure, izacard2023atlas}, fact-checking \cite{zeng2024justilm} through retrieval-augmented generation framework \cite{lewis2020retrieval}.
% When it comes to training LLM itself, 
\begin{figure}[t]
    % \vspace{-0.5cm}  % 调整与下文的间距
    \centering
    \includegraphics[width=0.5\textwidth]{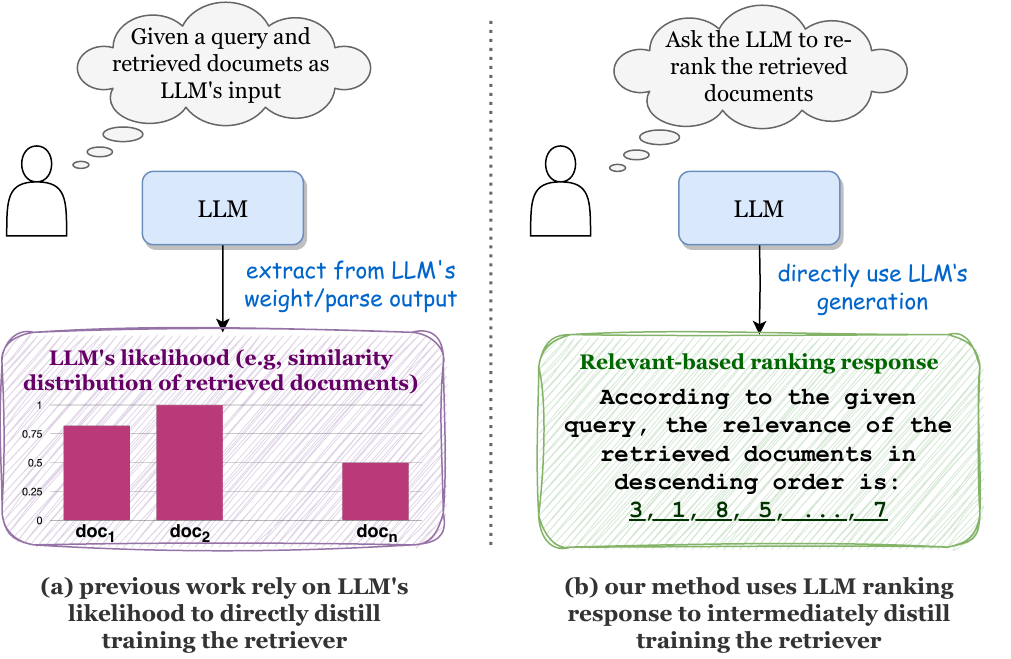} % Reduce the figure size so that it is slightly narrower than the column.
    \caption{Previous distillation methods (left) rely on extracting supervision signals from LLM's weights or using LLM's output probabilities to train the retriever model. 
    In contrast, our approach (right) bypasses the need for LLM's likelihood, directly using the LLM's ranking responses as supervision signals.}
    \label{fig:01}
    \vspace{-5mm}  % 调整与下文的间距
\end{figure}

The retrieval-augmented generation (RAG) framework has been widely adopted to alleviate hallucination problems in LLMs generation, especially for knowledge-intensive tasks \cite{lewis2020retrieval}. 
The RAG framework consists of two key components: a retriever to locate relevant information from a large corpus based on a given input, and a reader, typically a LLM, to integrate this information into its generation \cite{izacard2023atlas, shi2023replug}. 

% to enhance the content generation.
% Influenced by such impressive performance, there are also numerous studies which have explored how to effectively leverage the capabilities of LLMs to complement information retrieval models, leading to the development of joint-training in Retrieval-augmented Generation (RAG) framework \cite{lewis2020retrieval}. 

% During the inference stage, the information retrieved by the retriever improves the accuracy of the reader’s output and reduces the frequency of hallucination responses.

% traditional retrieval algorithms and 
% Moreover, unlike using the retrieval algorithms or off-the-shelf models, jointly training RAG further enhances the performance of the retriever from another perspective. Specifically, it uses the powerful semantic alignment capabilities of LLMs to distill and train the retrieval model, thereby improving the overall performance of the RAG framework.

How to distill knowledge from LLMs to optimize the retriever in the RAG framework with in-domain data has been a crucial challenge. 
Early efforts proposed training the retriever with white-box LLM readers by extracting supervision signals directly from the LLMs' weights \cite{izacard2023atlas, rubin2023long, guu2020retrieval}. 
However, this approach becomes more computationally intensive and time-consuming as LLMs increase in size.
Meanwhile, it is incompatible with closed-source models.

Recently, researchers have also turned to knowledge distillation for the retriever from black-box LLMs by training the retriever directly from generated outputs, such as RePLUG \cite{shi2023replug} and In-Context RALM \cite{ram2023context}. 
However, both methods use the generation log probabilities for correct answers as the distillation signal to train the retriever, which may suffer from: 
1) Limited application scenarios, as the output probabilities are not always available for closed-source LLMs. 
2) Discrepancy between retrieval and generation, where training LLMs' next-token prediction is not optimal for retriever training.
3) High computing costs, since hundreds of thousands of training instances are required in their training process.

To address these limitations, we propose \textit{Intermediate Distillation}, a data-efficient training scheme that leverages LLM-generated ranking responses to guide the training of the retriever. 
Our model employs a rerank-then-retrieve pipeline, where LLMs indirectly influence the retriever training via an intermediate ranker model.
We chose this pipeline for three main reasons:
% Specifically, we first leverage the generated reranking responses to distill a smaller re-ranker model. Based on this, we further utilize this well-trained re-ranker model to distillation training a small retriever model, thereby intermediately transferring part of the retrieval capabilities of the large language model to the smaller retriever. 
% Instead of direct distilling the retriever model, we use a rerank-then-retriever pipeline to let LLM intermediately guide the training of retriever model, and the reasons are mainly twofold. 
1) The robust zero-shot ranking capabilities of LLMs establish a strong foundation for knowledge distillation.
2) Using LLMs to generate a relevance-based ranking order is more suitable for retriever training than depending on LLMs output probabilities, making the supervision signals more reliable.
3) There are no restrictions on accessing this generated ranking order.
% the re-ranking capability of LLMs for retrieved documents according to its relevance to the query prompt is more reliable than their direct ability to scale the relevance of retrieved documents. 
% Compared to the former, generating natural language responses is much more challenging for the latter to provide more accurate answers for large language models.
% Second, as previous research has found, the learning ability of the small-scale retrievers is limited. Introducing a re-ranker model with stronger data fitting ability as a learning intermediary helps them better assimilate the knowledge from large language models. 

Specifically, we first train a ranker model using the ranking orders generated by LLMs as supervision signals. We then employ this trained ranker to further train the retriever model.
We conduct a series of experiments using advanced, closed-source LLMs that restrict output probability access.
The empirical results demonstrate the effectiveness of our method, requiring 100x to even 1000x less data than previous methods \cite{ram2023context, shi2023replug}, thereby significantly reducing computational costs.
Our main contributions are:
% \vspace{-20pt}
\begin{itemize}
    \setlength{\itemsep}{-2pt}
    \item We introduce Intermediate Distillation, a data-efficient knowledge distillation training scheme that optimizes retrieval models from black-box LLMs via an intermediate ranker model in a two-stage process.
    \item We conduct extensive experiments with cutting-edge LLMs and demonstrate the efficacy and efficiency of the proposed method in enhancing information retrieval performance compared to other supervision signals.
    \item We deploy our distilled retriever model within the RAG framework and demonstrate its effectiveness in downstream tasks such as open-domain question-answering.
   
\end{itemize}

% Meanwhile, the presence of the Retrieval-augmented Generation (RAG) framework \cite{lewis2020retrieval} has further enhanced the quality of LLMs' responses.
% That is, by leveraging the retriever model which dynamically searches the relevant information feed to the language model for generation, the corresponding responses become more accurate and the frequency of the hallucinations also reduces compared to using the vanilla language model.

% However, it has been a great challenge for the RAG framework to retrieve the passages that are most likely to be the target answers from a large amount of corpus, as training a high-quality retriever model is time-consuming and 

% Influenced by the large language models' impressive performance, previous works have explored the retrieve ability of LLM fine-tuning it as a retriever or a re-ranker, which shows the fine-tuned LLM achieves state-of-the-art performance, surpassing previous smaller retriever models. However, huge computational and time resources are needed to fine-tune a large language model before it becomes a high-quality retriever model. In this context, exploring how to utilize the capabilities of large language models more cost-effectively has become a valuable research question. 

\section{Related Work}
% Our proposed framework aims to utilizes the advanced semantic alignment capabilities of LLMs to distill their knowledge into smaller retriever models.
In this section, we provide a comprehensive background of information retrieval systems and knowledge distillation research related to LLMs.
% Since our proposed knowledge distillation framework aims to  the advanced semantic alignment capabilities of current large language models (LLMs) Therefore, we incorporate foundational concepts from information retrieval systems and knowledge distillation research to provide a comprehensive background.

\begin{figure*}
    \centering
    \includegraphics[width=1.0\textwidth]{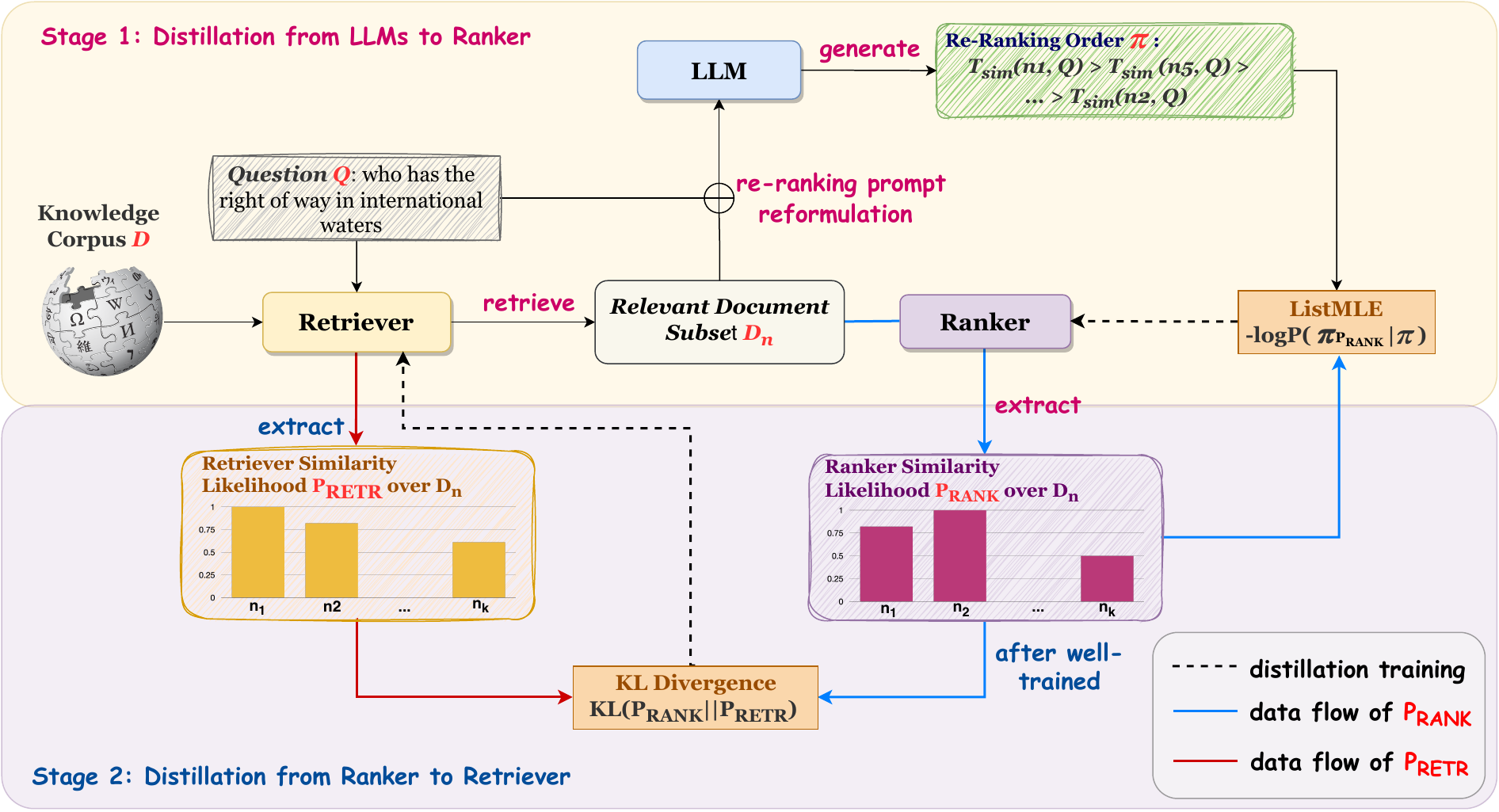}
    \caption{The two-stage knowledge distillation process of our proposed Intermediate Distillation scheme. In Stage 1, we use re-ranking order $\pi$ (highlighted in the green background color) as the supervisory signal to train a ranker model. In Stage 2, this distilled ranker unsupervised trains the retriever model to enhance its performance.}
    \label{fig:02}
\end{figure*}

\subsection{Retrieval-Augmented Generation}
Information retrieval plays a crucial role in various knowledge-intensive NLP tasks, including question-answering \cite{siriwardhana2023improving, zhang2024raft}, fact-verification \cite{hang2024trumorgpt, khaliq2024ragar} and open-domain dialogue \cite{wang2024unims, shuster2021retrieval}.
A prevalent approach in information retrieval is the multi-stage retrieval process \cite{nogueira2020document}, which first uses a retriever model to search several most relevant documents from the large corpus, then employs a ranker model to further optimize the ranking order based on relevance, and returns the top few most relevant documents finally.

% For example, researchers have fine-tuned LLMs to act as retrieval or re-ranking models, which outperform the existing smaller models. 
% Moreover, it has already demonstrated that LLMs are of excellent zero-shot reranking capabilities.
% The effectiveness of the RAG framework has been validated across various NLP downstream tasks, including open domain question-answering, fact-verification and dialogue.

Recently, the retriever models are increasingly used to enhance the generation quality of LLMs for knowledge-intensive tasks due to their flexibility and effectiveness, leading to the development of the retrieval-augmented generation (RAG) framework \cite{guu2020retrieval, izacard2023atlas}.
This framework integrates information retrieval into the generation process of LLMs, which helps overcome the models' limitations, such as hallucination, by utilizing external up-to-date information. 
% This method reduces the dependency of LLMs on their internal parameters to encode world knowledge, thereby reducing the occurrence of hallucination problems in LLMs generation.
% Normally, the RAG framework consists of a retriever model and a reader model. 
% The retriever is responsible for searching the relevant information from the large corpus, while the reader combines this retrieved information with query as input and generate responses.
In the RAG framework, the retrieved information can be in the form of tokens, entities, or text chunks (i.e., documents), and the retrieval can occur once or repeatedly every $n$ tokens, for finding a balance between the performance and time-cost.
% Specifically, the retrieved information form can be tokens, entities, or text chunks (i.e., documents); the retrieval process can be performed just once, or it can be executed every $n$ tokens, which could enhance the performance but also increase the inference time cost. 
Additionally, the retrieval model in RAG is adaptable to both encoder-to-decoder \cite{guu2020retrieval, izacard2023atlas} and decoder-only language models \cite{borgeaud2022improving, ram2023context}, and is applicable during both the pre-training \cite{zhong2022training, min2022nonparametric} and inference stages \cite{menick2022teaching, min2023factscore}. 

In this paper, we use the advanced knowledge from LLMs as the supervision signal to train the retriever models through a multi-stage (i.e., rerank-then-retrieve) training scheme.
We then integrate our well-trained retriever model into the RAG framework, 
demonstrating the effectiveness of our proposed training framework in question-answering tasks.
% demonstrating that our distillation method substantially improves the performance of LLMs in question-answering downstream tasks.

\subsection{Knowledge Distillation in LLMs.} 
Knowledge distillation is widely used to transfer knowledge from complex, large teacher models to smaller student models \cite{hinton2015distilling}.
Influenced by the outstanding performance of LLMs, more and more studies focus on using LLMs as teacher models to distill knowledge into smaller task-specific models \cite{brown-etal-2023-efficient}, and the distillation methods can be categorized into two types: white-box \cite{gu2023minillm, agarwal2023gkd, udagawa-etal-2023-comparative} and black-box \cite{li2022explanations, ho2022large, hsieh2023distilling}.
Specifically, white-box training leverages both the predictions and the parameters of LLMs to exact knowledge, which can be memory-intensive and computationally demanding. 
In contrast, black-box training only relies on the predictions of LLMs, making it less resource-intensive.

Many studies have successfully integrated knowledge distillation within the RAG framework to train the retriever models.
For white-box LLM distillation training, previous researches employ LLM likelihood, such as attention scores, to assess the relevance distribution of retrieved documents \cite{izacard2023atlas, izacard2022unsupervised}.
Meanwhile, some recent studies have also explored methods for training RAG using black-box LLMs, like In-Context RALM \cite{ram2023context} and RePLUG \cite{shi2023replug}.
% This kind of methods use the LLM's output probability distribution as the supervisory signal to train the 
% They directly transfer the semantic relevance of the retrieved information to the retriever model based on the next token's prediction probability, which is assumed to be the ground truth.
The remaining problem is that these methods still rely on the generation log probabilities for the ground truth as the supervision signals in distillation training, which tend to have a degree of randomness and are limited to the availability of the output probabilities.
Furthermore, aligning LLM predictions with the goals of retriever training is not the optimal choice since there remains a gap between retrieval and generation.
% Typically, using white-box training methods require deploying LLMs locally, which is complex to implement.
% Moreover, this kind of methods consumes increasingly more computational resources as the scale of LLMs models continues to expand, making it progressively more difficult to sustain. 
% Some recent studies have proposed methods for training Retrieval-Augmented Generation (RAG) based on black-box language models.
% Currently, there are also some studies that conduct training within the Retrieval-Augmented Generation (RAG) framework based on black-box LLMs, such as in-context learning and RePLUG. 
% Generally, these output probability-based methods have more randomness, and has limited scope since the output probabilities of LLMs are not always available.
% In addition, optimizing the prediction of LLM is not always matching the goal of training a retriever model, as there has a gap between retrieval and generation process.
% the training a retrieval model to semantically align directly with LLMs also requires extensive training data.
% These methods use LLMs to generate a series of quantitative score distributions to measure how much each retrieved document could improve the LM perplexity, thereby reveal the relevance of each document to the query. 
% However, this score distribution is not stable. 

In contrast, our proposed distillation method only requires black-box LLMs to output a relevance-based ranking order of the candidate relevant documents, yielding more consistent, matching, and interpretable results than output probability-based methods.
Moreover, our method is much more data efficient, requiring about 100x and 1000x less data compared to previous approaches \cite{shi2023replug, ram2023context}, significantly saving computational resources and increasing the flexibility of the training process.

\section{Method}

% We introduce Light LLM-Guider, a lightweight knowledge distillation method that treats LLM as a black-box and transfer its superior semantic alignment ability to reranker models and hence the performance of the retriever model within retrieval-augmented generation framework.

% 0112 delete
% Previous work in information retrieval often uses a ranker model to refine the order of information retrieved by the retriever model, aiming to improve the accuracy of the top results.
% In this work, we utilize the ranker model differently, using it as an intermediary for knowledge distillation from LLMs to the retriever model, as letting LLMs perform relevance-based ranking tasks for ranker training is simpler and more robust than having them directly generate a relevance likelihood for retriever training.

% we adapt this two-stage workflow, using a ranker model as an intermediary for knowledge distillation from LLMs to the retriever model instead of as a way to enhance the results' accuracy.
% As discussed in the previous sections, letting LLMs perform relevance-based ranking tasks is simpler and more robust than having them predict or directly generate a quantified likelihood of information relevance.

Our two-stage distillation scheme uses a ranker model and a retriever model as the student models and a LLM as the teacher model.
As shown in Figure \ref{fig:02}, we initially employ an off-the-shelf retriever to select a subset of documents $D_n$ from a large corpus $D$ based on their relevance to a query $Q$.
The LLM then re-ranks these documents, creating a ranking order $\pi$, which is used to train the ranker model in the distillation \textit{Stage 1}.
In \textit{Stage 2}, this ranker enhances the original retriever by minimizing the KL-divergence between their similarity likelihood.
% The following sections introduce our proposed framework in details, and is organized as follows.
In detail, Section \ref{sec:3-1} provides the formal definitions of the related tasks. 
In Section \ref{sec:3-2} and Section \ref{sec:3-3}, we show how knowledge is directly transferred from LLMs to a ranker model and then further conveyed to a retriever model, respectively.

\subsection{Problem Formulation}
\label{sec:3-1}
Given a question $Q$, the goal of a retriever model is to find a subset of the most relevant documents $D_n=\{n_1, n_2, ..., n_k\} \subseteq D$ from a large knowledge corpus $D=\{d_1, d_2, ..., d_m\}$, where each $d_i, n_i$ represents a unique document.
In the retrieval-augmented generation (RAG) framework, this subset $D_n$ is combined with the question $Q$ to form the reader's prompt input and then generates the corresponding answer $A$.

For the re-ranking task in our distillation framework, the teacher ranker model (i.e., LLMs) is tasked with reordering the documents $D_n$ according to their relevance to the question $Q$.
The re-ranking order can be represented as $\pi: T_{sim}(n_i, Q)>T_{sim}(n_k, Q) > ... > T_{sim}(n_j, Q)$, indicating the descending order of relevance from the documents $n_i, n_j, ..., n_k$ in the subset $D_n$.
This order information $\pi$ is first transferred to the ranker model, which serves as an intermediary between the LLM and the retriever model.
Subsequently, the ranker model conveys this knowledge to the retriever, thereby enhancing its performance.
% and serve as an intermediary to transfer the knowledge from LLMs to the retriever model finally.

\begin{figure*}
    \centering
    \includegraphics[width=1.0\textwidth]{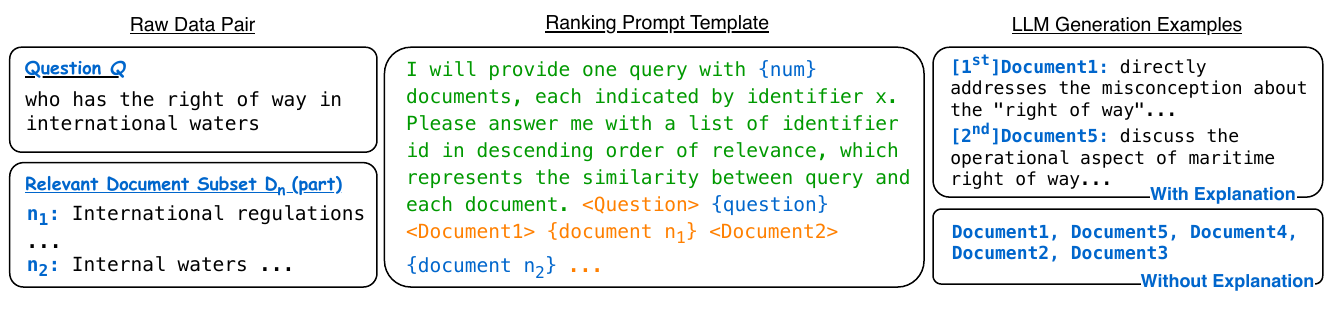}
    \caption{An example of our LLM teacher model's re-ranking process.}
    \label{fig:03}
    % \vspace{-3mm}
\end{figure*}

\subsection{Stage 1: Distillation from LLMs to Ranker} 
\label{sec:3-2}
% \textbf{Ranking Data Initialization} 
The initial step of our knowledge distillation workflow is data initialization, where we find the relevant document subsets $D_n$ from corpus $D$ for each specified question $Q$. 
These subsets then serve as the input for the \textit{Stage 1} training.
In practice, we employ a widely-used information retrieval model, Contriever \cite{izacard2022unsupervised}, as the retriever model for data initialization. \\
% Additionally, we initialize both the ranker and retriever models within our distillation framework using off-the-shelf Contriever model checkpoints.\\
\textbf{Re-ranking by LLMs.} 
In this stage, we utilize LLM's guaranteed zero-shot ranking capabilities to generate high-quality re-ranking orders for each subset $D_n$ based on the relevance to the corresponding question $Q$.
We use a list-wise ranking prompt, which is adapted from the RankGPT \cite{sun2023chatgpt}, as our input ranking prompt.
Figure \ref{fig:03} illustrates a LLM re-ranking process, where the LLM generates a re-ranking order $\pi_{D_n}: T_{sim}(n_1, Q)>T_{sim}(n_5, Q) >T_{sim}(n_4, Q) > T_{sim}(n_3, Q) > T_{sim}(n_2, Q)$. 
% We found that the LLM's excellent semantic alignment capabilities effectively pushed documents containing the corresponding answers to higher positions in the generated responses, as shown in Table 2 (where the numbers represent the average ranking position of documents containing the answers). 
Following this, we use these re-ranking orders to transfer LLM's knowledge into a smaller but more efficient ranker model.
 \\
\textbf{Ranker Distillation Training}
We initialize our ranker model by using the dual-encoder structure \textit{Contriever} checkpoint.
% To extract the ranker model's representation of the similarity between each document $n_i \subset D_n$
For each question $Q$ and its retrieved document $n_i \subseteq D_n$, we represent the question and document using the average value from the last hidden layer of the ranker model, denoted as $\hat{Q}$ and $\hat{n}_i$, respectively.
For each training data instance, once we have the representative embeddings $\hat{Q}$ and $\hat{n}_i$, we proceed to calculate the similarity likelihood $P_{RANK}$ over $D_n$, which can be defined as:
\begin{equation}
    P_{RANK}(n_i|Q) = \frac{exp(s(\hat{n}_i, \hat{Q})/\theta)}{\sum_{k=1}^Kexp(s(\hat{n}_k, \hat{Q})/\theta)}
\end{equation}
where $s$ denotes the dot-product between the ranker's representation vectors of the question and each retrieved document, and $\theta$ is the temperature hyper-parameter.
% \textbf{Ranker Training Objective} 

After obtaining the ranker model's similarity likelihood $P_{RANK}(n_i|Q)$ over its relevant document subset $D_n$, and the LLM-generated re-ranking order $\pi_{D_n}$, we use ListMLE \cite{xia2008listwise}, a list-wise loss function for distillation training from LLMs to the ranker model.
ListMLE considers the similarity likelihood $P_{RANK}(D_n)$ as the predicted list and $\pi_{D_n}$ as the ground truth, aiming to minimize the following loss function:
% where oi is the quality order of the i-th sentence measured by M . N1 is the number of sentences.
% \begin{equation}
%     L(f(x), y) = -log(P({\pi_{f(x)}}|y))
% \end{equation}
\begin{equation}
    \begin{aligned}
    L(P_{RANK}, \pi) 
    &= -log P(\pi_{P_{RANK}}|\pi) \\
    &= -log \prod _{j=1}^k \frac{exp(P_{RANK}(n_i))}{\sum_{m=j}^kexp(P_{RANK}(n_m))}
    \end{aligned}
\end{equation}
where $\pi_{P_{RANK}}$ represents the permutation of documents ordered by the ground truth ranking $\pi$. 
The loss function calculates an exponential probability distribution over all elements of $P(\pi_{P_{RANK}}|\pi)$, expressing the loss as the negative log-likelihood of the ground truth order $\pi$.
% and calculates the cumulative product across the sequence as defined by $\pi_{p_{RANK}}$.
% and $P({\pi_{f(x)}}|y)$ is the Plackett-Luce probability of a permutation $\pi_{f(x)}$ conditioned on scores $y$.
% Here we apply the exponential function to $P({\pi_{f(x)}}|y)$, which is:
% \begin{equation}
%     P({\pi_{f(x)}}|y) = \prod _{j=1}^n \frac{e^{f(x_i)}}{\sum_{k=j}^ne^{f(x_k)}}
% \end{equation}
% whose cumulative multiplication follows the permutation order $\pi$.
\subsection{Stage 2: Distillation from Ranker to Retriever}
\label{sec:3-3}
% After completing Stage 1 of our proposed framework, we obtain a well-trained ranker model. 
In \textit{Stage 2}, this well-train ranker model is used to enhance the retriever model's performance by transferring knowledge from LLMs.\\
% \textbf{Compute Retriever Similarity Likelihood.} 
We initialize our retriever model using a dual-encoder Contriever checkpoint, similar to the ranker.
For each question and its retrieved document, we also compute their representations $\widetilde{Q}$ and $\widetilde{n}_i$ by calculating the average value from the retriever model's last hidden layer and the similarity distribution $P_{RETR}$ of the retriever model over $D_n$ is defined as:
\begin{equation}
    P_{RETR}(n_i|Q) = \frac{exp(s(\widetilde{n}_i, \widetilde{Q})/\theta)}{\sum_{k=1}^Kexp(s(\widetilde{n}_k, \widetilde{Q})/\theta)}
\end{equation}
where $s$ represents the dot-product between the retriever's representation vectors of the question and each retrieved document, and $\theta$ is the temperature hyper-parameter.

% \textbf{Retriever Training Objective} 
We then leverage the similarity likelihood $P_{RANK}$ from the previously trained ranker model to enhance the retriever model's performance under an unsupervised learning process by minimizing the KL-divergence between $P_{RANK}$ and $P_{RETR}$:
\begin{equation}
    D_{KL}(P_{RANK}||P_{RETR})
\end{equation}

This process ensures the retriever model aligns more closely with the text similarity knowledge from the LLMs.
Through this two-stage distillation scheme, we enhance the retrieval accuracy and effectiveness of the retriever model, which can be further applied to the RAG framework to improve its performance on knowledge-intensive NLP tasks.

\section{Experiment}
\label{sec:4}
\setlength{\tabcolsep}{1.5mm}{
\begin{table*}[t]
    \centering
    \scalebox{0.9}{
    \resizebox{1.0\textwidth}{!}{
    \begin{tabular}{*{10}{c}}
      \toprule
      \multirow{2}*{\textbf{Distillation Methods}} & \multicolumn{4}{c}{\textbf{NQ}} & \multicolumn{4}{c}{\textbf{TriviaQA}} & \\
      \cmidrule(lr){2-5} \cmidrule(lr){6-10}&
      \textbf{HR@5$\uparrow$} & \textbf{HR@10$\uparrow$} & \textbf{EM$\uparrow$} & \textbf{F1$\uparrow$} & & 
      \textbf{HR@5$\uparrow$} & \textbf{HR@10$\uparrow$} & \textbf{EM$\uparrow$} & \textbf{F1$\uparrow$}
      \\
      % \midrule
      % w/o RAG & --- & --- & 19.39 & 29.21 & & --- & --- & 51.64 & 59.47 \\
      \midrule
      w/o Distillation & 0.478 & 0.583 & 26.09 & 36.75 &  & 0.595 & 0.678 & 54.99 & 63.55 \\
      % \midrule
      % \rowcolor{gray!9}
      % \multicolumn{10}{c}
      % {\textbf{\textsl{Direct Distilation}}}\\
      % \midrule
      % GPT-4o & 0.505 & 0.617 & 26.23 & 36.47 &  & 0.623 & 0.699 & 55.39 & 63.96 \\
      \midrule
      \rowcolor{gray!9} 
      \multicolumn{10}{c}
      {\textbf{\textsl{Supervised Distillation}}}\\
      % \textit{Compared Supervised Signal} & & & & & & & & & \\
      \midrule
      BM25 & 0.186 & 0.262 & 18.17 & 27.88 & & 0.120 & 0.175 & 46.90 & 55.38 \\
      Rule-Based & 0.223 & 0.303 & 19.75 & 29.64 & & 0.277 & 0.356 & 50.08 & 58.45 \\
      Metric (ROUGE-2) & 0.534 & 0.643 & 27.76 & 38.16 &  & 0.641 & 0.716 & 56.17 & 64.92 \\
      \midrule
      \rowcolor{gray!9} 
      \multicolumn{10}{c}
      {\textbf{\textsl{Intermediate Distillation (Ours)}}}\\
      \midrule
      GPT-3.5 Turbo & 0.505 & 0.606  & 25.84 & 36.13 & & 0.587 & 0.664 & 53.72 & 62.16 \\
      GPT-4o & 0.553 & 0.652 & 27.01 & 37.38 &  & 0.664 & \textbf{0.734} & 56.27 & 64.98 \\
      GPT-4 Turbo & 0.545 & 0.656 & 28.31 & 38.68 & & 0.662 & 0.727 & 56.15 & 65.07 \\
      Claude3 Opus & \textbf{0.562} & \textbf{0.665} & \textbf{28.45} & \textbf{38.83} & & \textbf{0.669} & 0.733 & \textbf{56.68} & \textbf{65.36} \\
      % Atlas-large & 38.51 & 47.42 & 0.655 &\\
      \bottomrule
    \end{tabular}
    }}
    \caption{The performance comparison of our proposed Intermediate Distillation scheme with other baseline supervised distillation methods on question-answering tasks.}
    \label{tab:tab02}
    \vspace{-5mm}  % 调整与下文的间距
\end{table*}}
\subsection{Experiment Setup}
\textbf{Dataset} We conduct experiments on two benchmark open-domain question-answering datasets: \textit{NaturalQuestions (NQ)} \cite{kwiatkowski-etal-2019-natural} and \textit{TriviaQA} \cite{joshi-etal-2017-triviaqa}.
The \textit{NQ} dataset includes queries from \textit{google.com query} and their corresponding Wikipedia pages, each with an annotated passage containing the answer.
We use the dataset version provided by ATLAS \cite{izacard2023atlas} and follow its training, validation, and testing splits: 79,168/8,757/3,610. 
% Similarly, the document source of the question-answer pairs in \textit{TriviaQA} dataset is also collected from Wikipedia and the web. 
Similarly, we also use the \textit{TriviaQA}, which contains question-answer pairs sourced from Wikipedia and the web, that ATLAS provides and following its training, validation, and testing splits: 78,785/8,837/11,313.
For the knowledge corpus base, we utilize data from
Wikipedia as of December 20, 2018, adapting the passage embeddings provided by ATLAS.

For our experiments, we selectively sample 1,000 instances from each training set from \textit{NQ} and \textit{TriviaQA}, keeping validation and testing sets unchanged.
We further discuss the impact of selecting different types of training data in Section \ref{sec:5-2}.
This training set size is about 100 to 1,000 times smaller than those used in previous black-box LLM distillation methods within the RAG framework, demonstrating the superior data efficiency of our approach. 
% We sample only 1,000 data instances from the original training set partitions of the \textit{NaturalQuestions} \textit{TriviaQA} datasets to form our final training dataset while keeping the size of validation and test sets unchanged.
% The training data sets we select consist of question-answer pairs where the corresponding relevant retrieved document subset contains the answer (i.e., ground truth) but is not directly ranked first in the subset.
% We choose this training set size and select strategy for two reasons: 1) Obtaining responses from closed-source LLMs leads to considerable costs, and using 1,000 entries instead of tens of thousands of entries strikes a balance between economic expenses and improvements in model performance; 2) Our experimental results, which detailed in the following section, show that this sampling strategy significantly improves the retriever model's performance within the RAG framework, outperforming other training methods.
The impact of training set size on performance is discussed further in Section \ref{sec:5-3}. \\
% Section 5 further explore how training set size affects the retriever model's performance improvement.\\
% \textbf{Experimental Settings} 
% We use the pre-trained \textit{Contriever} model as our initial reranker model and retriever model. 
% For knowledge distillation training, we choose the high-performing GPT3.5, GPT4o, GPT4-turbo and Claude3 to serve as the teacher model.\\
\textbf{Baseline} We evaluate our distillation framework, Intermediate Distillation, against the following established text similarity methods: \textit{ROUGE-2}, an evaluation metric frequently used in NLP, and \textit{BM25}, a popular information retrieval algorithm.
The idea of employing NLP evaluation metrics like ROUGE-2 for knowledge distillation in retriever models is first proposed by \cite{he2022metric}, which also uses a multi-step distillation approach to solve the Commonsense Reasoning tasks \footnote{We choose ROUGE-2 as our compared baseline metric, as it outperforms other metrics in this prior study.}.
For both ROUGE-2 and BM25, we use the similarity likelihood between the query and its relevant documents via their calculation to generate re-ranking order as the supervision signals.
Meanwhile, we do not consider the previous work \textit{RePLUG} \cite{shi2023replug} as a baseline since it uses a larger data scale and relies on LLMs output probabilities, which is not a fair comparison.

In addition, we conduct a \textit{Rule-Based} experiment that ranks documents containing the answer at the top in re-ranking order, which aims to demonstrate that effective re-ranking distillation signals should not only highlight answers.\\
% We evaluate the effectiveness of our methods and baseline approaches on retriever models, as well as the retriever's performance with the RAG framework.\\
\textbf{Experimental Settings} 
We initialize our ranker and retriever models using the Contriever checkpoint \cite{izacard2022unsupervised} with a dual-encoder structure.
For our proposed distillation scheme, we select several cutting-edge and representative LLMs as the teacher models, including GPT-3.5 Turbo, GPT-4o, GPT-4 Turbo, and Claude3 Opus.
To comprehensively evaluate the performance of our retriever model, we integrate the distilled retriever model into the RAG framework for question-answering tasks, which allows us to measure the improvement in the quality of responses generated by the language model. 
In the RAG framework, we use the reader model Llama-3-8B-Instruct \cite{touvron2023llama} to generate answers.
We also evaluate a baseline version of this RAG framework without additional distillation training for the retriever (i.e., \textit{w/o Distillation} experiment). \\
\textbf{Implantation Details} We set both the ranker and retriever models with a hidden layer size of 768, thus totally have approximately 10 million training parameters for each model.
The learning rates are set as 5e-5 for the ranker model and 2e-5 for the retriever model. 
Both models are trained for 5 epochs on the \textit{NQ} and \textit{TriviaQA} datasets, using a batch size of 20 and optimized with the Adam optimizer.
Additionally, we restrict the size of the relevant document subset $D_n$ to 5, each retrieved document with a maximum length of 128. We further discuss the impact of the retrieve subset (i.e., re-ranking list) size in Section \ref{sec:5-4}.
\\
\textbf{Evaluation Metrics} We evaluate the retrieval performance of our distilled retriever model through the \textit{top-5} and \textit{top-10} retrieval Hit Rates (HR@5 and HR@10), which is the percentage of questions where the relevant document subset $D_n$ includes at lease one correct answers with the top-5 and top-10 documents.
For question-answering tasks in the RAG framework, we use the standard Exact Match (EM) metric and F1-Score to evaluate the accuracy and precision of the language model generated responses.

\subsection{Experimental Results}
We present our experimental results, including all the baseline methods and settings evaluated on the testing set of \textit{NQ} and \textit{TriviaQA} in Table \ref{tab:tab02}\footnote{The highest values in the table are highlighted in bold on both the NQ and TriviaQA datasets.}.
The experimental results show that the retriever model, under our proposed Intermediate Distillation scheme and supervised by Claude3, achieves the best performance in most evaluation metrics, confirming the effectiveness of our proposed method.

Moreover, the quality of supervision signals from LLMs greatly influences the performance of the distilled retriever models.
For example, the retriever model trained under GPT-4 Turbo supervision outperforms the one supervised by GPT-3.5 Turbo within our Intermediate Distillation scheme, aligning with GPT-4 Turbo’s higher performance across various NLP tasks \cite{openai2023gpt4}. 
As the rapid development of LLMs, this improvement in supervision quality has also evolved: from being less effective than the ROUGE-2 metric (i.e., as seen the retriever under supervised by GPT-3.5 Turbo) to significantly surpass it (i.e., supervised by GPT-4 Turbo).

In the RAG framework for question-answering tasks, a stronger retriever model is more likely to enhance the output quality of the reader model, demonstrating the effectiveness and adaptability of the Intermediate Distillation framework for NLP downstream tasks.
% At the same time, while the trend between the retriever's performance and the quality of the reader's output in RAG usually aligned, they do not always match perfectly.
However, according to our experimental results, while Intermediate Distillation using GPT-4o typically outperforms Supervised Distillation using ROUGE-2 in retrieval performance, the latter can still produce higher quality generations within RAG. 
This divergence may be due to the different objectives of the retriever and the reader: the retriever focuses on accurately identifying the ground truth, whereas the reader wants the retriever to provide information that more effectively helps the reader in generating accurate responses.
This discrepancy remains a topic that can be further explored in future work.

\section{Analysis}
In this section, we conduct ablation studies and a series of quantitative analyses to evaluate how various experimental designs and settings affect the outcomes of our distillation results.
\label{sec:5}
\subsection{Ablation Studies}
\label{sec:5-1}
In this subsection, we conduct an experiment named \textit{Direct Distillation}, where we train the retriever model directly using the relevance likelihood generated by LLM. More details of experiment setting can be found in Appendix \ref{sec:appendix-A}.
We compare the results of this approach with our proposed two-stage distillation scheme under the same LLM teacher model (i.e., GPT-4o), and the experimental results are shown in Table \ref{tab:tab03}.
The results indicate that the Direct Distillation method is less effective than our proposed Intermediate Distillation scheme, which further validates the rationality of the two-stage design of our proposed framework. 
% GPT-4o & 0.505 & 0.617 & 26.23 & 36.47 &  & 0.623 & 0.699 & 55.39 & 63.96 \\
% \vspace{-2mm}
\subsection{Impact of the Training Data Type} 
\label{sec:5-2}
\setlength{\tabcolsep}{1.5mm}{
\begin{table}[t]\label{tab:performance-drp}
    \centering
    \scalebox{1.0}{
    \resizebox{0.5\textwidth}{!}{
    \begin{tabular}{*{6}{c}}
      \toprule
      \multirow{3}*{\textbf{Method}} & \multirow{3}*{\textbf{Dataset}} & \multicolumn{3}{c}{\textbf{Evaluation Metrics}} & \\
      \cmidrule(lr){3-6}
      & & \textbf{EM$\uparrow$} & \textbf{F1$\uparrow$} & \textbf{HR@5$\uparrow$} \\
      \midrule
      Direct Distillation & NQ & 26.23 & 36.47 & 0.505 \\
      & TriviaQA & 55.39 & 63.96 & 0.623 \\
      \midrule
      Intermediate Distillation & NQ & 27.01 & 37.38 & 0.553 & \\
      & TriviaQA & 56.27 & 64.98 & 0.664 \\
      \bottomrule
    \end{tabular}
    }}
    \caption{Ablation studies on the effectiveness of two-stage distillation scheme design.}
    \label{tab:tab03}
    \vspace{-5mm}  % 调整与下文的间距
\end{table}}
\begin{figure}
    \centering
    \includegraphics[width=0.45\textwidth]{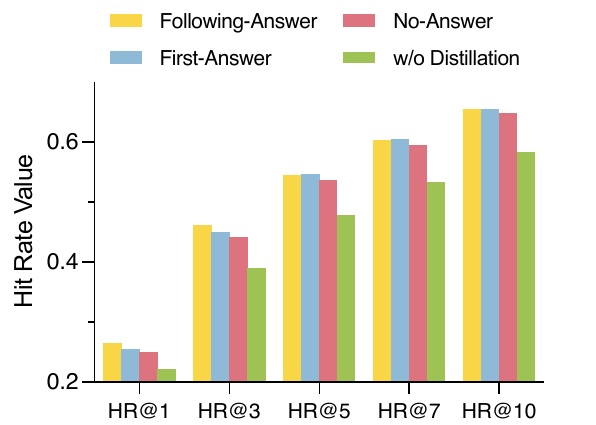}
    \caption{The performance of retriever models across three types of training sets, which vary based on the initial appearance and placement of ground truth in the retrieved subsets.}
    \label{fig:04}
    \vspace{-10pt}
\end{figure}
Our previous experiment demonstrate that Rule-Based supervision signals, which places documents containing answers at the top, are ineffective and detrimentally impacting the retriever's performance.
This indicates that simply re-ranking documents based solely on the presence of ground truth (i.e., the correct answer) does not provide the high-quality text similarity insights required for effective distillation.
To delve deeper into the influence of the appearance and placement of ground truth in the re-ranking process, we categorize the initial retrieved document subsets $D_n$ based on \textit{NQ}'s queries into three categories:
(1) \textit{Following-Answer}: contains at least one document with the correct answer, but this kind of document is not at the first position in the subset. 
This data type is used in our experiments detailed in Section \ref{sec:4}.
(2) \textit{First-Answer}: contains at least one document with the correct answer, and this kind of documents is at the first position in the subset.
(3) \textit{No-Answer}: no documents in the subset contain the correct answer.

We follow the same training setting used in our primary experiments in Section \ref{sec:4}, and use GPT-4 Turbo as the LLM teacher model.
% The performance of the retriever models under these three data types of training are shown in Figure \ref{fig:04}.
% The results indicate that subsets with having ground-truth documents (i.e., Following-GT and First-GT) significantly improve retriever performance more than those without (i.e., No-GT).
% distilling the retriever model using subsets of relevant documents that contain the ground truth (i.e., Follow-GT and First-GT) produces better results compared to using the types that do not contain the answer (i.e., No-GT), and using Follow-GT proves to be more effective than using First-GT overall. 
Together with findings from the \textit{Rule-Based} experiments in Section \ref{sec:4}, the experiment results shown in Figure \ref{fig:04} indicate that considering the semantic similarity of the text is far more important than arranging documents containing the answers to the top for re-ranking in distillation training, as even the retriever under the \textit{No-Answer} data set training has notable improvements.

Moreover, as the \textit{Following-Answer} training data, where the correct answers are not ranked first initially, yields better training results than using the the \textit{First-Answer} training data, indicating that optimizing the ground truth placement in re-ranking also has a positive effect on the experimental results after the consideration of text similarity.
% Combined with the results from the Rule-Based experiments, this experimental results show that both prioritizing documents containing the answers and considering the semantic similarity of the text beyond just the answers determine the quality of the re-ranking distillation signals.

% This results suggests that re-ranking information containing the ground truth, especially those not initially ranked first, is a more effective distillation method. 
% Moreover, even the retriever under the No-GT data set training has notable improvements, showing that which highlights the advanced ranking capabilities of LLMs.
% \vspace{-1mm}
\subsection{Impact of the Training Set Size} 
\label{sec:5-3}
Our previous experiments demonstrate that our distillation framework significantly enhances retriever model performance with just 1,000 training instances. 
In this subsection, we explore how different training set sizes affect distillation effectiveness.
We use training sets of of 50, 100, 200, 500, 1000, and 2000 data instances from the \textit{Following-Answer} data type, with other settings consistent with our experiments in Section \ref{sec:4}. 
In addition, we use GPT-4 Turbo as our LLM teacher model.
% In addition to this, we use the same training setting as Section \ref{sec:5-1} uses to train the retriever model, and the experimental results are shown in Figure \ref{fig:05}.

\begin{figure}
    \centering
    \includegraphics[width=0.5\textwidth]{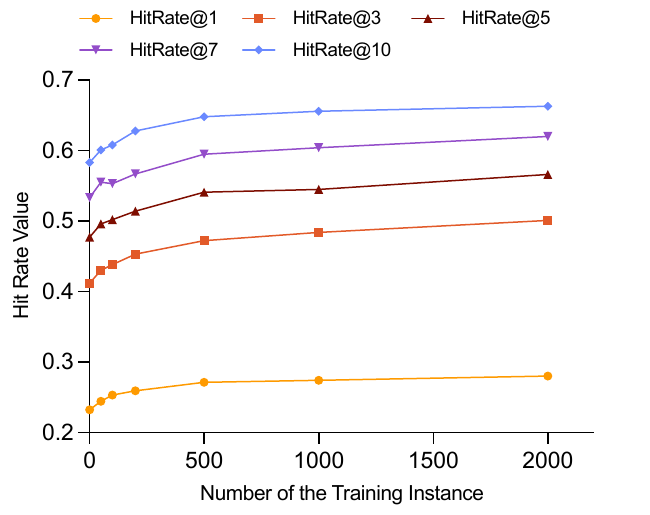}
    \caption{The performance of retriever models under different training set size.}
    \label{fig:05}
    \vspace{-4mm}
\end{figure}

Results in Figure \ref{fig:05} show that the performance of the retriever model improves significantly with training data with thousands of or even only hundreds of instances.
These empirical findings highlight the data efficiency of our proposed distillation scheme.
% initial increases in performance are significant with small training sets. 
In addition, although initial performance increases are notable with small training sets, the rate of improvement decreases as more training data is used.
This pattern indicates a \textit{scaling law} in distillation training, where further enhancements become increasingly difficult as the model's performance improves.
For models that already perform well, even marginal improvements require much more data, demanding greater training resources.

% For already high-performing retriever models, even small improvements require an exponentially larger amount of training data.
\subsection{Impact of the Re-ranking List Size}
\label{sec:5-4}
\begin{figure}[!htbp]
    \centering
    \includegraphics[width=0.5\textwidth]{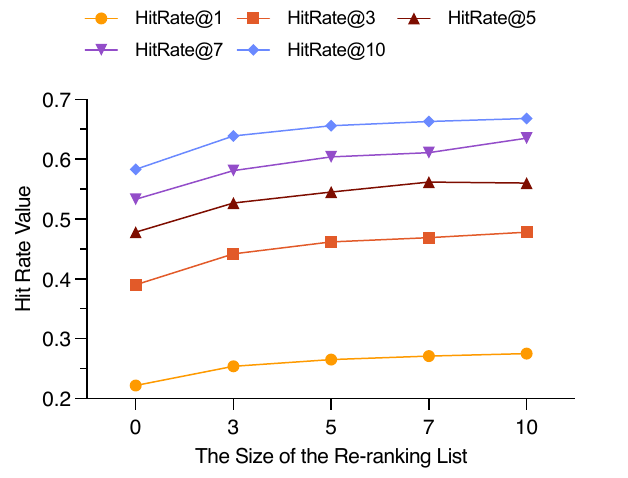}
    \caption{The performance of retriever models under different size of the re-ranking list. The performance corresponding to 0 re-ranking list size represents the baseline retriever model performance.}
    \label{fig:06}
    \vspace{-4mm}
\end{figure}
In previous experiments, we set the re-ranking list to five documents (i.e., we retrieve five relevant documents each time).
Generally, larger re-ranking lists offer more supervision signals from LLMs, thus potentially enhancing the effectiveness of distillation training. 
To explore the impact of re-ranking list size on our distillation method, we vary the re-ranking list sizes, using the top-3, top-5, top-7, and top-10 documents from each relevant retrieved subset to conduct the distillation training.
% That is, during the initial data collection phase, the off-the-shelf retriever model returns the top-3, top-5, top-7, and top-10 relevant documents, respectively.
We keep other training settings consistent with those in our primary experiments in Section \ref{sec:4} and use GPT-4 Turbo as the LLM teacher model.

The experimental results shown in Figure \ref{fig:06} show that increasing the re-ranking list size progressively improves the effectiveness of the distillation training. 
As the list expands from re-ranking three documents to ten documents, the performance of the distilled retriever model consistently improves.
Moreover, compared with the retriever model's baseline performance, setting the size of the re-ranking list to 3 still significantly improves the retriever model's performance not only in HitRate@3 but also across broader metrics from HitRate@5 to HitRate@10.

% TODO in appendix:
% direct distill prompt cases
% case studies in metric v.s. LLM RAG performance

\section{Conclusion}
We propose Intermediate Distillation, a two-stage data-efficient knowledge distillation scheme that uses the remarkable capabilities of black-box LLMs to train an information retrieval model through an intermediate ranker model.
We conduct extensive experiments with advanced LLMs, demonstrating that our method enhances the effectiveness and efficiency of the retriever model performance compared to other supervision signals. 

\section{Limitation}
This paper proposes a data-efficient distillation scheme using black-box LLMs to train smaller information retrieval models, which prove its effectiveness with training data on the scale of thousands. 
However, we do not evaluate our proposed distillation scheme with larger scales of training data, such as tens of thousands or millions of instances, due to budget limitations on accessing responses from closed-source LLMs and insufficient computational resources to utilize high-quality open-source LLMs like Llama-70B. 
In the future work, we will focus on extending this study to larger-scale training data, using either closed-source or advanced open-source LLMs to further analysis the effectiveness of our proposed distillation scheme.

\bibliography{custom}

\appendix

\clearpage
\section{Analysis of Different Distillation Signal Generated by LLMs}
\label{sec:appendix-A}
Here we provide more details about the prompt design of the \textit{Direct Distillation} experiment.
We also analyze the corresponding generation quality of LLMs, including the stability and interpretability, and compare it with the re-ranking generation.

\subsection{Direct Distillation Prompt Design}
We follow the re-ranking prompt format design and replace the re-ranking task with quantifying the similarity scores of the retrieved documents for LLM's supervised signals generation. An example input prompt and the generated responses without explanations is as follows:\\
\newline
\underline{Input Prompt}:
\begin{verbatim}
I will provide one query with {num} 
documents, each indicated by number 
identifier x. 
Please answer me with a list of the
similarity score between the provided 
query and documents based on your judgment. 
The score should be between 0-1. 
Please don’t use interoperator and only 
output the score list.
<Question> {question} 
<Document1> {document n1} 
...
\end{verbatim}
\underline{Generated Response}:
\begin{verbatim}
[0.1, 0.1, ...., 0.8, 0.2]
\end{verbatim}

\subsection{Comparison with Re-ranking Response}
We compare the responses generated from re-ranking prompts and those derived from similarity score prompts used in the Direct Distillation experiment. 
Specifically, we randomly select 1,000 data instances from the \textit{NQ} dataset, using the queries and their retrieved documents to prompt the LLM to generate both list-wise re-ranking orders and similarity scores.
Examples of responses generated by the LLM, specifically using GPT-4 Turbo, are shown in Table \ref{table:03}.

\begin{figure}[!htbp]
    \centering
    \includegraphics[width=0.48\textwidth]{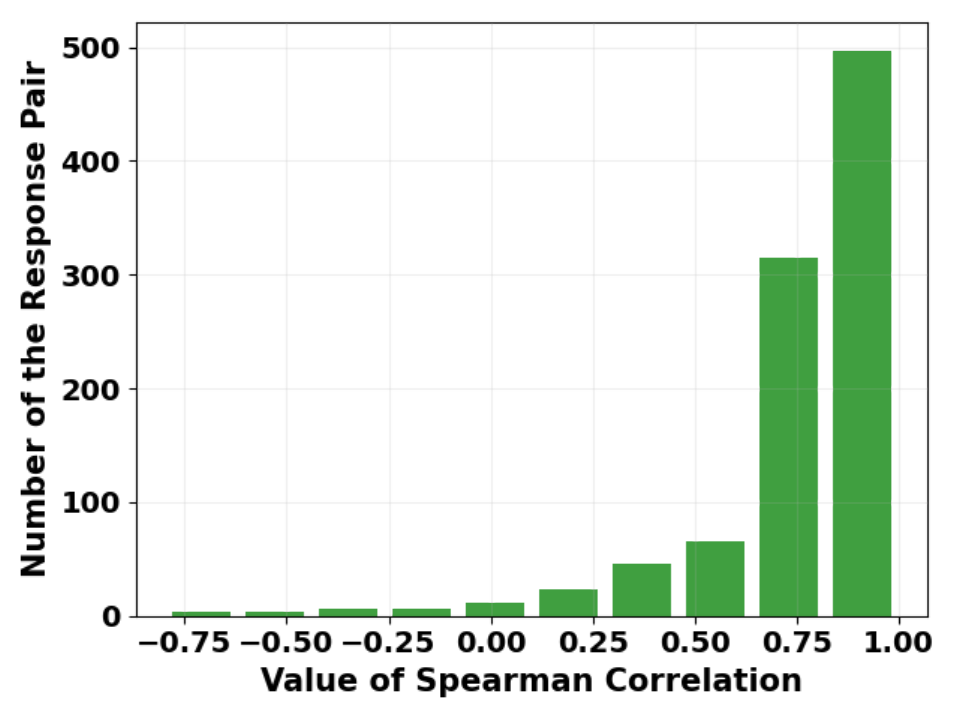}
    \caption{Spearman Correlation between the responses from the re-ranking prompt and the responses from the similarity score prompt.}
    \label{fig:07}
    \vspace{-4mm}
\end{figure}

Table \ref{table:03} shows that responses generated from re-ranking prompts are more interpretable than those from similarity score prompts, which often use scaled values that are ambiguous.
Additionally, responses from the similarity score-based prompt frequently yield extreme values, such as 0.0 (completely dissimilar) or 1.0 (highly similar), in some cases, which means that  that supervision signals based on similarity scores are less informative and act more like binary signals in certain data instances.

Moreover, we use the \textit{Spearman Correlation} to assess the consistency between responses generated from re-ranking prompts and those from similarity score prompts, and the analysis result is visualized in Figure \ref{fig:07}.
A higher Spearman correlation value suggests a stronger positive correlation between the two types of responses.
From Figure \ref{fig:07}, we can see that many response pairs are closely related, indicating the stability and reliability of LLMs in generating responses for similar tasks.
In addition, based on our previous analysis, we can see that the responses from re-ranking prompts are not only reliable but also possess a higher information density compared to those from similarity score prompts, showing that responses from re-ranking prompts have higher-quality supervision capabilities.

\subsection{Implantation Details of Direct Distillation Experiment}
We use the same training settings as those used in Intermediate Distillation experiments.
Superficially, we set the retriever model with a hidden layer size to 768 and initialize it using a dual-encoder Contriever checkpoint.
The model is trained over five epochs using the same dataset as described in Section \ref{sec:4}, with a learning rate of 5e-5, a batch size of 20, and optimization via the Adam optimizer.
In addition, we also limit the size of the relevant document subset $D_n$ to five, and each retrieved document with a maximum length of 128.

\begin{table*}[!ht]
\small
\centering
\begin{tabular}{|p{6cm}|c|p{6cm}|}
\hline
\textbf{Question \& Retrieved Documents} & \textbf{Type} & \multicolumn{1}{c|}{\textbf{LLM's Generation}} \\ 
\hline
\multirow{2}{=}
{\textbf{[Question]} membership in the european union requires countries to have which type of government
\textbf{[Document1] } the european continent. the eu's example is being followed by the african union, the union of south American nations...
\textbf{[Document2] } powers of states, the considerable level of translation and the choice of a lesser city as the capital...
\textbf{[Document3]} principal characteristic of classical {\color{red}democracy}. the concept of a mixed government was studied during the renaissance and the age of reason by niccolo machiavelli, giambattista vico...
\textbf{[Document4]} thus essentially preventing it from participating in government. the european union ( eu ) is an association of twenty - eight sovereign member states...
\textbf{[Document5]} civiletti opinions, not all funding gaps led to shutdowns. of the nine funding gaps between 1980 and 1990, only four led to furloughs...} & \textit{Re-ranking}
& [\textit{Re-ranking order}] {\color{blue}Document5, Document3, Document2, Document1, Document4}.
\textit{[Explanation]} Document5 directly discusses the governmental systems used by most European nations, particularly the parliamentary system, which is relevant since the EU generally comprises nations with parliamentary democracies. This makes it the most relevant to the query about the type of government required for EU membership. Document3 talks about mixed constitutions...\\
\cline{2-3}
& \textit{Similarity Score} & 
[\textit{Similarity scores from Document1 to Document5}] {\color{blue}Doc1:0.0, Doc2:0.1, Doc3:0.3, Doc4:0.1, Doc5:0.7}. 
\textit{[Explanation]} Document1 talks about the EU and various other international organizations, focusing on aspects like the single market and policy integration but does not mention government types required for EU membership. Hence, the score is 0.0. Document2 discusses the structure of EU governance... \\
\hline

\multirow{2}{=}
{{\textbf{[Question]} when was the immigration reform and control act passed}
\textbf{[Document1] } united states temporarily for money or a better life. an increase in legal immigrants migrating to the united states also led to an increase of illegal aliens...
\textbf{[Document2] } there are noted communists, socialists, and anti - american sympathizers. parts of the act remain in place today...
\textbf{[Document3] } ...the immigration reform and control act ( irca ), also known as the simpson 2013 mazzoli act or the reagan amnesty, signed into law by ronald reagan on {\color{red}november 6}, 1986...
\textbf{[Document4] } five percent of the u. s. workforce. it was also estimated that about 70 percent of those illegal workers...
\textbf{[Document5] } to reduce them as restrictionists. "fry traced" restrictions roots " to a surprising surge in illegal and legal immigration...
}
& \textit{Re-ranking} & [\textit{Re-ranking order}] {\color{blue}Document3, Document1, Document4, Document5, Document2}.
\textit{[Explanation]} Document3 directly answers the query by explicitly mentioning the "Immigration Reform and Control Act of 1986" and stating it was signed into law by Ronald Reagan on November 6, 1986. It provides the most specific information related to the query... \\
\cline{2-3}
& \textit{Similarity Score} & [\textit{Similarity scores from Document1 to Document5}] {\color{blue}Doc1:0.0, Doc2:0.0, Doc3:1.0, Doc4:0.0, Doc5:0.0}. 
\textit{[Explanation]} Document1 discusses general immigration trends and issues but does not provide the specific date of the passage of the Immigration Reform and Control Act, which is directly asked in the query. Therefore, it receives a score of 0.0. Although Document2 mentions other immigration-related legislation and acts, it does not specify...\\
\hline

\multirow{2}{=}
{\textbf{[Question]} who sings in there's something about mary  
\textbf{[Document1] } melody. mary described the completed song as "whitney - fied". meron maintains that...
\textbf{[Document2] } ...he film concludes with the two engaging in a kiss while singer / songwriter {\color{red}jonathan richman}, who has periodically narrated the story in song throughout the film...
\textbf{[Document3] } day but her. minutes later bing notices his tackle box has been disturbed, and an instant later the kids see the door to the tent is open...
\textbf{[Document4] } true opinions of other contestants and her stance on the judges. the book achieves in combining fifty years of a'salt of the earth'existence with a new - found life of fame...
\textbf{[Document5] } and breezy number and the narrator opens up the story about bing and his family. and while bing is cleaning the fish, he tells the story of " goldilocks"...
}
& 
\textit{Re-ranking} & [\textit{Re-ranking order}] {\color{blue}Document2, Document5, Document4, Document1, Document3}.
\textit{[Explanation]} Document2 is the most relevant sentence as it directly answers the query. It mentions that Jonathan Richman narrated parts of the story through song throughout the film "There's Something About Mary," which clearly identifies the singer related to your query... \\
\cline{2-3}
& \textit{Similarity Score} & [\textit{Similarity scores from Document1 to Document5}] {\color{blue}Doc1:0.0, Doc2:1.0, Doc3:0.0, Doc4:0.0, Doc5:0.0}. 
\textit{[Explanation]} Document1 discusses a song in a completely different context, related to "Cinderella" and Whitney Houston's music adjustments, not related to the film "There's Something About Mary." Thus, it doesn't answer the query about who sings in the specified film, receives a score of 0.0. Document2 directly answers the query. It mentions that Jonathan Richman narrated parts of the story... \\
\hline    
\end{tabular}
\caption{Examples of two different LLM-generated responses. We also let LLM generate the corresponding explanations in these examples. The red color indicates the corresponding answer to the question.}
\label{table:03}
\end{table*}

% \section{Case Studies of Metric and LLM Distillation in RAG}

\end{document}